\documentclass[pdflatex,sn-mathphys-num]{sn-jnl}% Math and Physical Sciences Numbered Reference Style
\usepackage{graphicx}%
\usepackage{multirow}%
\usepackage{amsmath,amssymb,amsfonts}%
\usepackage{amsthm}%
\usepackage{mathrsfs}%
\usepackage[title]{appendix}%
\usepackage{xcolor}%
\usepackage{textcomp}%
\usepackage{manyfoot}%
\usepackage{booktabs}%
\usepackage{algorithm}%
\usepackage{algorithmicx}%
\usepackage{algpseudocode}%
\usepackage{listings}%
\theoremstyle{thmstyleone}%
\theoremstyle{thmstyletwo}%

\theoremstyle{thmstylethree}%

\begin{document}

\title[Article Title]{Mid-infrared single-pixel imaging via two-photon optical encoding}


\author[1]{\fnm{Huijie} \sur{Ma}}

\author*[1,2,3]{\fnm{Kun} \sur{Huang}}\email{khuang@lps.ecnu.edu.cn}

\author[1,2]{\fnm{Jianan} \sur{Fang}}

\author[1]{\fnm{Ziyu} \sur{He}}

\author[4]{\fnm{Yan} \sur{Liang}}

\author*[1,2,5,6]{\fnm{Heping} \sur{Zeng}}\email{hpzeng@phy.ecnu.edu.cn}

\affil[1]{\orgdiv{State Key Laboratory of Precision Spectroscopy, and Hainan Institute}, \orgname{East China Normal University}, \orgaddress{\city{Shanghai}, \postcode{200062}, \country{China}}}

\affil[2]{\orgdiv{Chongqing Key Laboratory of Precision Optics}, \orgname{Chongqing Institute of East China Normal University}, \orgaddress{\city{Chongqing}, \postcode{401121}, \country{China}}}

\affil[3]{\orgdiv{Collaborative Innovation Center of Extreme Optics}, \orgname{Shanxi University}, \orgaddress{\city{Taiyuan}, \postcode{030006}, \country{China}}}

\affil[4]{\orgdiv{School of Optical Electrical and Computer Engineering}, \orgname{University of Shanghai for Science and Technology}, \orgaddress{\city{Shanghai}, \postcode{200093}, \country{China}}}

\affil[5]{\orgdiv{Shanghai Research Center for Quantum Sciences}, \orgaddress{\city{Shanghai}, \postcode{201315}, \country{China}}}

\affil[6]{\orgdiv{Chongqing Institute for Brain and Intelligence}, \orgname{Guangyang Bay Laboratory}, \orgaddress{\city{Chongqing}, \postcode{400064}, \country{China}}}


\abstract{Mid-infrared (MIR) imaging offers powerful capabilities for label-free chemical analysis, yet its practical deployment remains hindered by the high cost and cryogenic complexity of conventional cameras. Two-photon absorption (TPA) provides a promising route to room-temperature MIR detection, but existing TPA imagers based on raster scanning or array detectors are constrained by slow acquisition speed or limited detection sensitivity. Here we present a scanning-free MIR single-pixel imaging approach based on non-degenerate TPA in a silicon detector. The involved spatial encoding is realized by a near-infrared structured pump with a resolution of 7 $\mu$m, thus allowing high-fidelity MIR optical modulation through the phase-matching-free nonlinear interaction. Consequently, the spatially modulated TPA response is intrinsically integrated in the single-element photodetector, which favors computational reconstruction of the impinging MIR image by correlating measured intensities and predetermined patterns. Notably, the use of advanced algorithms of compressed sensing and deep learning facilitate image recovery under sub-Nyquist sampling with a compression ratio of 10\% and photon-starved illumination with an incident light flux of 0.5 pJ/pulse. Furthermore, a multispectral imaging over 2.5-3.8 $\mu$m is manifested for chemical discrimination of plastic films. The presented architecture would offer a broadband and sensitive alternative for MIR imaging in various fields ranging from biomedical diagnostics to material inspection.}




\maketitle

\section*{Introduction}

The mid-infrared (MIR) spectral region provides powerful molecular fingerprinting capabilities, enabling precise identification of vibrational and rotational transitions associated with specific chemical bonds \cite{Cheng2015Science}. Nowadays, MIR imaging has become indispensable across diverse fields, including biomedical diagnostics, food safety inspection, remote sensing, and infrared surveillance \cite{Vodopyanov2020Wiley, Shi2020NM, Israelsen2019LSA}. To date, conventional MIR detectors are typically based on narrow-bandgap semiconductors such as mercury cadmium telluride (MCT) and indium antimonide (InSb), which suffer from high dark currents and limited sensitivity, particularly under room-temperature operation \cite{Rogalski2011IPT}. While cryogenic cooling can significantly enhance the performance, such as in superconducting nanowire detectors capable of detecting single infrared photon over a wide spectral range \cite{Taylor2023Optica}, it introduces substantial system complexity and power overhead. In parallel, emerging MIR detectors based on low-dimensional materials such as graphene \cite{Guo2018NM},  black phosphorus \cite{Bullock2018NP}, and colloidal quantum dots \cite{Xue2023LSA}, have demonstrated great potential for elevating the operating temperature. However, these devices still confront critical challenges, including low broadband responsivity, limited operational stability, and challenges in large-area fabrication \cite{Wang2019Small}.

In contrast, silicon-based detectors for visible and near-infrared (NIR) wavelengths exhibit high quantum efficiency and low dark noise, even enabling single-photon sensitivity at room temperature \cite{Hadfield2009NP}. Leveraging these advantages, indirect detection strategies have been developed to establish the so-called frequency upconversion detection, where MIR signals are transduced into the visible/NIR domain \cite{Barh2019AOP}. A widely adopted approach utilizes nonlinear media to perform wavelength conversion, either through quantum spectral correlations \cite{Kviatkovsky2020SA, Paterova2020SA, Cai2024SA} or optical parametric processes \cite{Dam2012NP, Huang2022NC, Ge2023PRAppl, Rehain2020NC}. Alternatively, the intrinsic nonlinearity of semiconductor detectors can be harnessed for MIR sensing via two-photon absorption (TPA) \cite{Boitier2009APL, Fang2020PRA} or higher-order multiphoton processes \cite{Pearl2008APL, Nevet2011OL}, offering distinct advantages to eliminate phase-matching constraints and simplify optical alignment. Particularly, the non-degenerate TPA (ND-2PA) configuration enable broadband MIR detection, which extends the operational window into the far-infrared while simultaneously enhancing nonlinear absorption coefficients \cite{Fishman2011NP, Boiko2017APL, Piccardo2018APL}. Moreover, recent efforts have demonstrated the feasibility of TPA imaging by exploiting the superiority of Si or InGaAs cameras, like high pixel density, large spatial formats, and high frame rates \cite{Knez2020Light, Liu2022IEEE, Knez2022SA}. Nevertheless, wide-field TPA imaging typically demands a large pump beam to spatially overlap with the MIR signal, which inevitably results in reduced pump intensity and degraded nonlinear efficiency \cite{Fang2021IEEE}. While raster-scanning approaches can provide high sensitivity across an extended field of view \cite{Pattanaik2016OE}, the inherently slow acquisition speed limits practical applicability. These limitations highlight the ongoing need to realize scanning-free, wide-field MIR TPA imaging with high detection sensitivity.

In this context, single-pixel imaging paradigm offers a compelling solution by combining a single-point detector with spatial encoding, thereby eliminating the need for expensive focal plane array detectors while delivering high sensitivity and rapid temporal response \cite{Edgar2019NP, Zhang2015NC, Kilcullen2022NC, Meng2024LSA}. Furthermore, the compatibility with compressed sensing and deep learning algorithms allows enhancement in both sampling efficiency and image reconstruction fidelity \cite{Duarte2008IEEE, Bian2017JOSAB, Zhang2020NC}. Currently, most commercially available spatial light modulators are optimized for the visible or NIR spectral ranges, thereby limiting their applicability in the MIR regime. Although digital micromirror devices (DMDs) can be customized to support MIR operation by resorting to specialized window materials \cite{Ebner2023SR}, parasitic diffraction effects significantly degrade the spatial modulation precision, especially at longer infrared wavelengths \cite{Edgar2019NP}. Notably, MIR modulators based on graphene metasurfaces have recently been demonstrated \cite{Zeng2018LSA, Lan2024LPR}, albeit with low pixel density and limited modulation efficiency. Therefore, it remains challenging to realize high-resolution MIR single-pixel TPA imaging, which calls for novel techniques to realize precise spatial modulation and sensitive photon detection.

Here, we devise and implement a single-pixel MIR imaging system based on nonlinear structured detection via the ND-2PA effect. Specifically, the MIR image is relayed onto a single-element silicon photodiode pumped by a NIR optical field with dynamic spatial patterns. The involved two-photon optical encoding not only facilitates a high-precision and high-speed spatial modulation with a spatial resolution of 7 $\mu$m, but also favors high-sensitivity and wide-field upconversion detection through the phase-matching-free nonlinear process. Consequently, the optically pumped large-bandgap semiconductor serves as both the modulator and detector for the MIR radiation, which contrasts to conventional single-pixel imagers with separate modulation and detection units. Intrinsically, the TPA response of the masked MIR signal is spatially integrated on the single-element sensor, delivering a sequence of intensity values corresponding to the programmed spatial patterns. Thanks to the advanced algorithms of compressed sensing and convolutional neural networks, high-contrast MIR images can be efficiently reconstructed under photon-starved conditions with an incident MIR pulse energy as low as 0.5 pJ and a sampling ratio of only 10\%. Furthermore, MIR spectral imaging over 2.5-3.8 $\mu$m is demonstrated to identify visually transparent plastic thin films. The presented novel configuration features single-pixel simplicity, high spatial resolution, and scanning-free operation, which would offer a promising alternative to broadband and sensitive MIR imaging at room temperature.

\section*{Methods}
\subsection*{Basic principle}
\begin{figure*}[b!]
	\centering
	\includegraphics[width=0.95\textwidth]{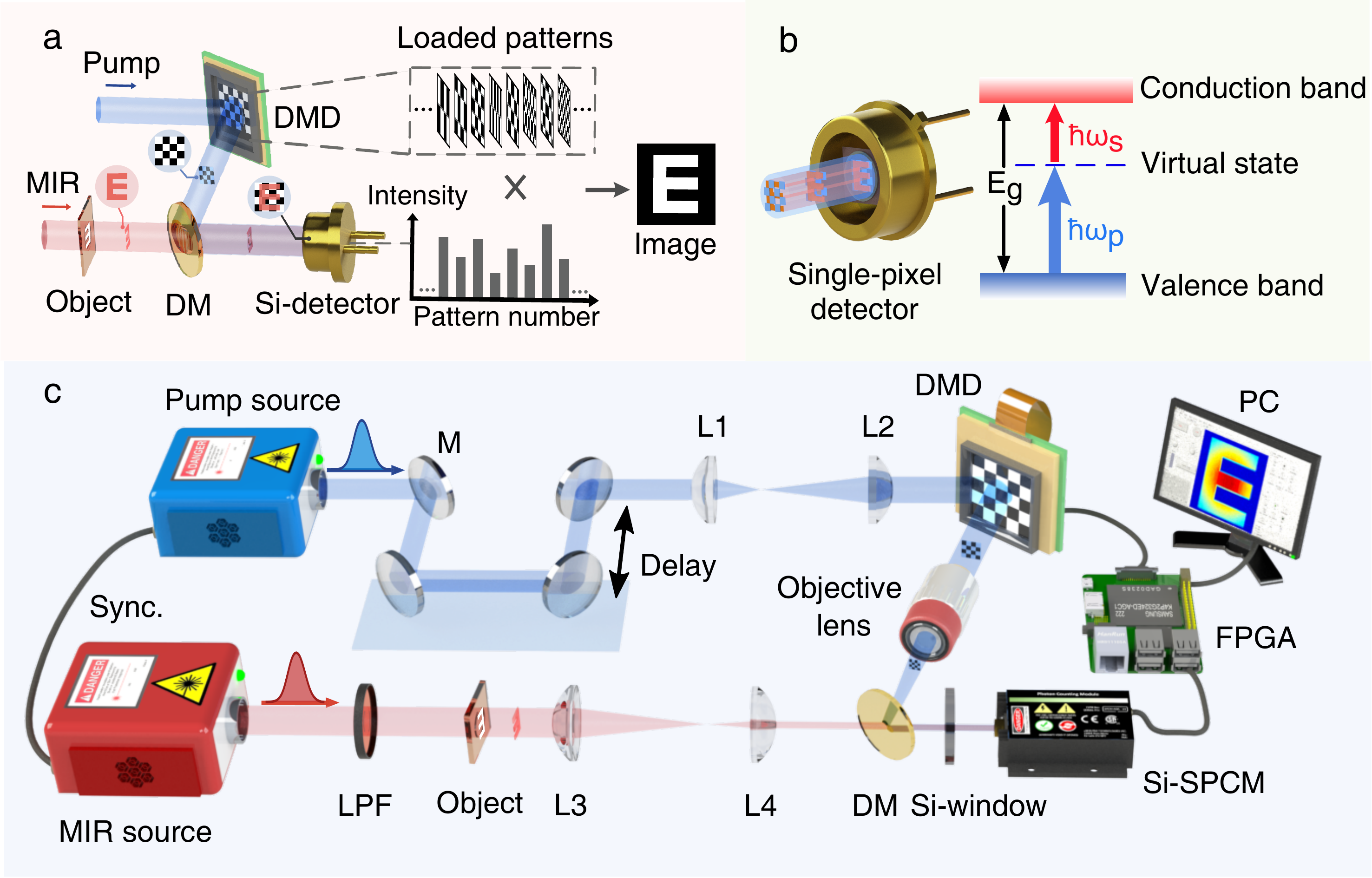}
	\caption{Schematic of MIR single-pixel ND-2PA imaging. (a) Conceptual illustration of the MIR single-pixel imaging approach. The MIR image on a single-element silicon sensor is optically modulated by a sequence of NIR structured pump patterns. The measured ND-2PA intensity is used to reconstruct the targeted image. (b) Energy-level diagram of the ND-2PA process, where the combined photon energy of the signal and pump fields is larger than the semiconductor bandgap ($E_g$). (c) Experimental setup. MIR illumination at around 3 $\mu$m passes through a transmissive mask patterned with a letter ``E", and then is relayed onto a silicon single-photon counting module (Si-SPCM) via a 4f imaging system. Meanwhile, a synchronized pump pulse at 1.55 $\mu$m is spatially modulated via a digital micromirror device (DMD), before being projected onto the detector with an objective lens. The resulting ND-2PA signal corresponds to the spatial integration of the optically modulated MIR beam. Finally, the measured intensity values along with the encoding patterns allow computational image recovery. Note that a silicon window as a spectral filter is used to block ambient noises. The involved timing control and data acquisition are managed by a field-programmable gate array (FPGA). L: lens; LPF: long-pass filter; M: silver mirror; DM: dichroic mirror.}	
	\label{fig1}
\end{figure*}

Figure \ref{fig1}(a) illustrates the concept of the operation principle for the MIR single-pixel TPA imaging, where the structured NIR pump imposes a spatial modulation onto the MIR image on a silicon photodiode. The all-optical approach allows high-fidelity MIR manipulation by using mature NIR spatial light modulators \cite{Stantchev2020NC, Wang2023NC}. Further combination of upconversion process facilitates the nonlinear structured detection, leading to the recent realization MIR single-pixel imaging at the single-photon level \cite{Wang2023NC}. Nevertheless, the upconversion process is performed in a nonlinear crystal, resulting in severe constrain on the field of view due to the stringent phase-matching requirement. In contrast, our scheme leverages the intrinsic TPA effect of the detector. The involved nonlinear process is free of phase matching, which depends solely on the electronic structure and nonlinear susceptibility of the absorbing medium, instead of the coherence length or phase velocity differences between optical fields \cite{Fishman2011NP, Knez2020Light}. 

As shown in Fig. \ref{fig1}(b), the silicon material simultaneously absorbs signal and pump photons with energies of $\hbar \omega_s$ and $\hbar \omega_p$, respectively. When the combined photon energy exceeds the semiconductor bandgap, $\textit{i.e.},$ $\hbar \omega_s + \hbar \omega_p > E_g$, electronic excitation from the valence band to the conduction band occurs to generate a measurable electrical photocurrent. When the incident photons have identical energies, the process is referred to be degenerate TPA (D-2PA). In comparison, ND-2PA exhibits significantly higher absorption coefficients due to resonance effects for the involved intermediate virtual states during the nonlinear photon-matter interaction \cite{Ziemkiewicz2024JCP}, which favors high-sensitivity MIR detection. Specifically, the generated ND-2PA signal follows
\begin{equation}
	I_{\text{ND-2PA}} \propto I_s \times I_p \ ,
	\label{eq1}
\end{equation}
where $I_{s,p}$ denote the signal and pump intensity. This relationship implies that spatial modulation of the MIR field can be equivalently achieved by encoding the NIR pump beam. The structured NIR illumination effectively serves as an all-optical mask for the MIR signal. To experimentally verify this nonlinear spatial mapping, we compare the NIR pump patterns captured by an InGaAs CCD with the corresponding ND-2PA response recorded by a silicon CCD under uniform MIR illumination. As shown in Supplementary Note 1, the two sets of patterns exhibit high spatial consistency, confirming the high fidelity of the modulation transfer.

Moreover, the ND-2PA response of the masked MIR signal is intrinsically integrated over the area of the detector sensor, as expressed by 
\begin{equation}
	P_\text{ND-2PA}(x, y) =\iint I_{\text{ND-2PA}} (x, y) dxdy \ .
	\label{eq2}
\end{equation}
The detected power by a single-element detector is directly correlated to each optical pump pattern. Given a sequence of orthonormal patterns $I_P^{(n)}(x,y)$, where $n$ is the pattern sequence number, the corresponding differential intensity signals between the positive and inverse patterns $P_\text{ND-2PA}^{(n)}$ are measured to provide weighted coefficients in the subsequent image reconstruction. Based on $N$ patterns, the two-dimensional image of the object $\mathcal{O}(x, y)$ can be reconstructed by
\begin{equation}
	\mathcal{O}(x,y) = \frac{1}{N} \cdot \sum_{n=1}^{N} P_\text{ND-2PA}^{(n)} I_P^{(n)}(x,y) \ .
	\label{eq3}
\end{equation}
Therefore, the nonlinear structured pumping allows simultaneous functionalities of all-optical spatial encoding and two-photon upconversion detection, which lays the foundation to implement the MIR single-pixel TPA imaging with a silicon detector.

\subsection*{Experimental setup}
Figure \ref{fig1}(c) illustrates the experimental setup for MIR single-pixel imaging based on the structured-pumping ND-2PA. The initial light source originates from a Yb-doped fiber laser (YDFL, LangyanTech, YbFemto ProH) at 1030 nm, which delvers mode-locked pulses at a repetition rate of 20.1 MHz. The pulse duration is measured to be about 1 ps by an optical auto-correlator. A portion of YDFL output is used to prepare the pump source based on an optical parametric amplifier (OPA). In the OPA, a continuous-wave light at 1550 nm from an extended-cavity diode laser is injected to a periodically poled lithium niobate (PPLN) crystal. Under the phase-matching condition with a poling period of 30.3 $\mu$m and a operation temperature of $48\,^\circ\mathrm{C}$, a train of ultrashort pulses at 1550 nm can be generated as the NIR pump. The other part of the YDFL output is injected into a fiber-feedback optical parametric oscillator (FOPO), allowing the generation of temporally synchronized MIR pulses \cite{Yu2024PR} as the MIR signal. The FOPO features a low pumping threshold and a wide tunable range of 2.5 to 3.8 $\mu$m. The passive synchronization configuration not only enables high-peak-power coincidence pumping to enhance the nonlinear conversion efficiency in the ND-2PA process, but also establishes a picosecond-scale temporal gate that effectively rejects continuous, incoherent ambient thermal noise, thereby improving the detection sensitivity of the MIR signal.

Then, the MIR beam after being filtered by a 2.4 $\mu$m long-pass filter is directed onto an object engraved with letters. The transmitted image is projected onto a silicon-based single-photon counting module (SPCM, Laser Components, SAP500T6) via a 4f relay imaging system. The SPCM is specified with a a spectral response range of 400-1100 nm and an active photosensitive diameter of 500 $\mu$m. Meanwhile, the pump beam is expanded to illuminate a DMD (Texas Instruments, V-650L) to perform the  structured spatial modulation. The modulated light is reflected via a dichroic mirror into the detector. A microscope objective is employed to facilitate a high-fidelity mapping, which has a focal length of 45 mm and a numerical aperture of 0.15. The scaling factor of the optical mask can be flexibly adjusted by adapting the distances of the involved elements, thus allowing a full cover of the MIR image while maintaining a high spatial resolution of the dynamic mask. In front of the SPCM, a silicon window (LBTEX, OW2-Si) is placed to eliminate ambient noises to which the detector is linearly responsive. Moreover, the ND-2PA response is optimized by adjusting the temporal overlap between the signal and pump pulses via the optical delay line. Finally, the measured sequence of TPA signals corresponding to the programmed patterns enable the recovery of targeted image. The involved data acquisition and timing control are managed by a field-programmable gate array (FPGA) \cite{Wang2023NC}. We note that precise spatiotemporal alignment between the MIR signal and the structured NIR pump is crucial for ensuring efficient TPA process and high-fidelity image reconstruction. This is achieved through passive synchronization and fine delay tuning for temporal overlap, as well as carefully designed relay optics and iterative beam alignment for spatial overlap. Together, these measures ensure stable and optimal nonlinear interaction at the detector plane.

\section*{Results and discussion}
\subsection*{TPA responses on a single-element detector}

\begin{figure*}[b!]
	\centering
	\includegraphics[width=0.9\textwidth]{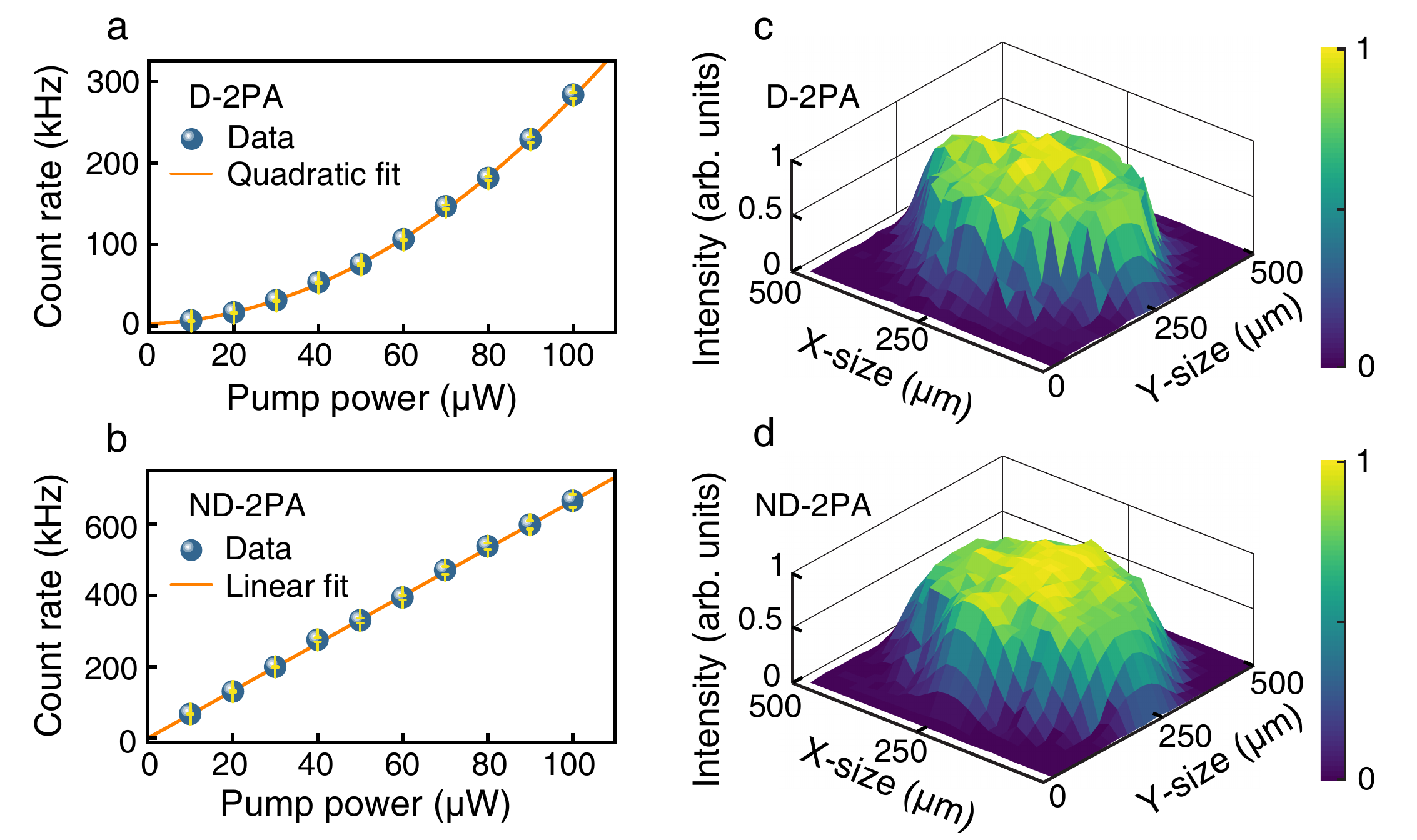}
	\caption{Characterization of TPA responses of the single-pixel silicon detector. (a) Measured count rate of the D-2PA signal versus the pump power. (b) Recorded ND-2PA signal count rate versus the pump power, with the MIR pulse energy fixed at 30 pJ. (c, d) Spatial distribution for the D-2PA (c) and ND-2PA (d) responses on the surface of the single-element sensor.}
	\label{fig2}
\end{figure*}

We first quantitatively characterize the nonlinear optical response of the single-element photon detector. Figure \ref{fig2}(a) presents the D-2PA response as a function of the pump power, indicating a quadratic behavior as expected. With the presence of MIR signal at 2700 nm, the recorded ND-2PA signal shows a linear dependence on the pump power, as shown in Fig. \ref{fig2}(b). Note that the MIR signal power is fixed at 30 pJ/pulse, the contribution due to the degenerate three-photon absorption (D-3PA) is negligible as discussed in Supplementary Note 2. Actually, the measured ND-2PA signal also exhibits a linear dependence on the MIR power, as governed by Eq. \eqref{eq1}. The bilinear response of the ND-2PA to both the pump and signal is important to ensure a rigorous foundation for the nonlinear optical modulation.

In our proposed single-pixel imaging scheme, the ND-2PA signal is intrinsically integrated within the detector, in contrast to previous approaches where spatial modulation and detection are implemented as separate components \cite{Stantchev2020NC, Wang2023NC}. As a result, the spatial uniformity of the detector's nonlinear response is critical for maintaining high-fidelity optical encoding. To quantitatively assess this uniformity, both the MIR and NIR beams are tightly focused onto the SPCM, enabling high-resolution spatial probing under constant pump and signal intensities. The detector is mounted on a high-precision two-dimensional translation stage to perform raster scanning across the entire active area. As shown in Figs. \ref{fig2}(c) and (d), the measured D-2PA and ND-2PA responses exhibit excellent spatial uniformity, confirming the consistency of the nonlinear detection across the sensor surface.

\subsection*{Single-pixel imaging via structured pumping}	
%\subsection{MIR single-pixel TPA imaging via structured optical pumping}

\begin{figure*}[b!]
	\centering
	\includegraphics[width=0.9\textwidth]{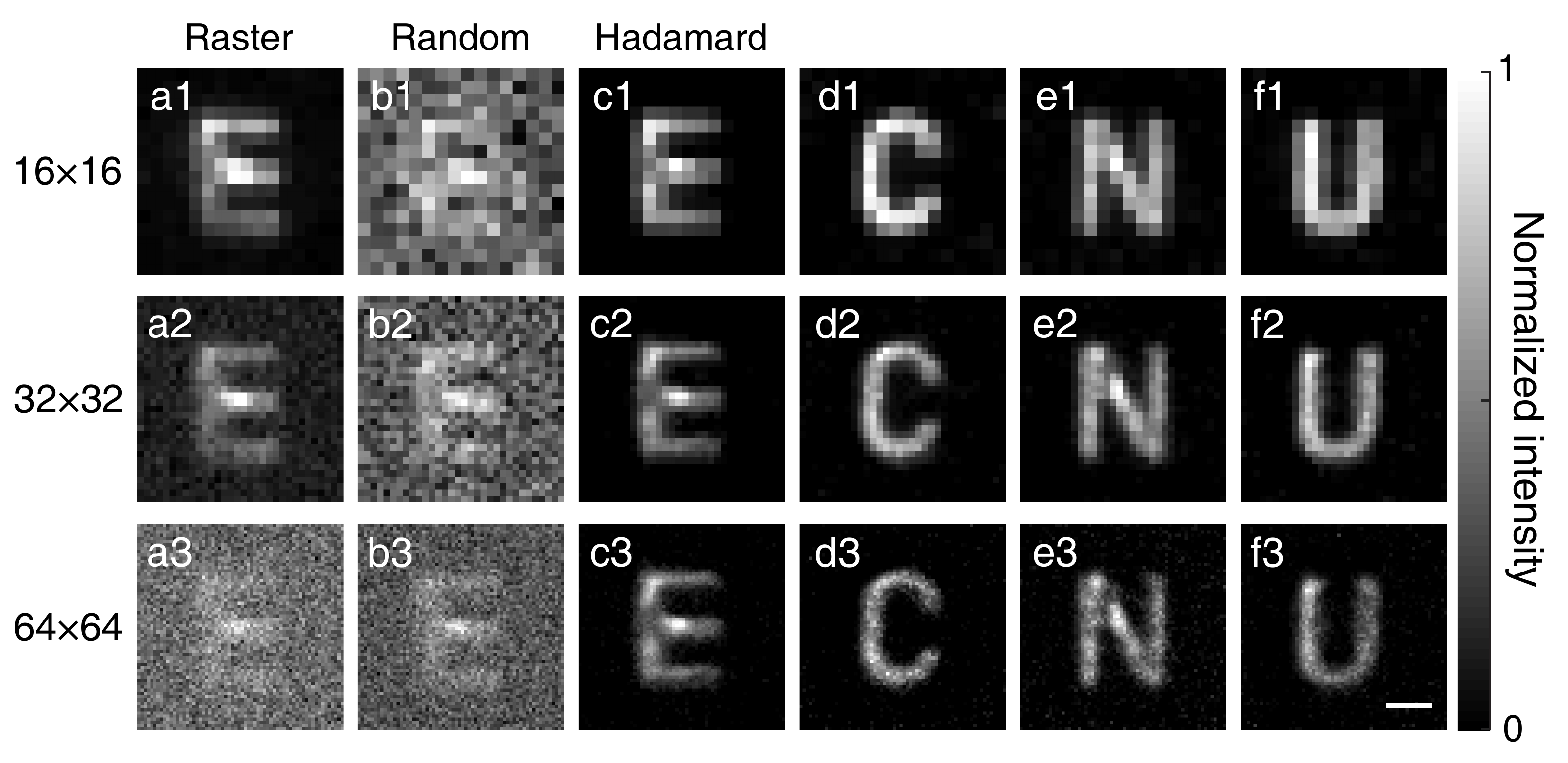}
	\caption{Performance comparison of reconstructed images under various encoding matrices. (a1-a3) Measured images in the case of using raster-scanning operation with increasing pixel numbers from 16$\times$16, 32$\times$32, and 64$\times$64, respectively. (b1-b3, c1-c3) Reconstructed images with multipixel masks based on Random (b1-b3) and Hadamard (c1-c3) patterns with increasing number of pixels from 16$\times$16, 32$\times$32 and 64$\times$64, respectively. (d1-d3, e1-e3, f1-f3) Reconstructed images in the case of Hadamard encoding for three other objects with letters ``C" (d1-d3), ``N" (e1-e3), and ``U" (f1-f3), respectively. Note that all the images are acquired at the same MIR illumination power. The scale bar corresponds to 100 $\mu$m.}
	\label{fig3}
\end{figure*}
Next, we turn to evaluate the performance of MIR single-pixel TPA imaging system based on the structured optical pumping.  In the single-pixel imaging paradigm, the choice of encoding matrix plays a pivotal role in determining reconstruction quality and noise resilience \cite{Edgar2019NP, Bian2017JOSAB}. To benchmark the imaging performance, we compare three types of encoding masks. The most straightforward method involves sequentially illuminating individual pixels, which effectively performs a raster scanning across the target scene. This per-pixel approach performs adequately under high light levels, where sufficient signal-to-noise ratio (SNR) can be maintained, as illustrated in Fig. \ref{fig3}(a1). However, raster scanning is inefficient in utilizing the available light, making it highly susceptible to detector noise. This limitation becomes especially pronounced at higher spatial resolutions, where the smaller physical size of each pixel substantially reduces the photon flux collected within a fixed integration time. As the signal level approaches the detector's noise floor, the signal-to-noise ratio deteriorates significantly. This degradation is clearly observed in Fig. \ref{fig3}(a3), where the 64$\times$64 resolution yields poor image fidelity, and the target object becomes barely discernible.

In contrast, multi-aperture masking schemes mitigate the detector noise by distributing light across multiple pixels during each measurement, thereby increasing the photon budget per acquisition. To implement this, the DMD is used to generate binary spatial illumination masks. Reconstructed images using random binary encoding are shown in Figs. \ref{fig3}(b1-b3) for 16$\times$16, 32$\times$32 and 64$\times$64 resolutions, respectively. Compared to raster scanning, random encoding offers noticeably improved SNR for a smaller pixel aperture where per-pixel photon counts are limited. A more effective encoding strategy utilizes Hadamard matrices, which form an orthonormal basis set that minimizes the mean squared error per pixel, thereby enhancing reconstruction quality \cite{Bian2017JOSAB}. In addition, the Hadamard encoding facilitates efficient reconstruction using the fast matrix transpose operation \cite{Wang2023NC}. In this work, we adopt the cake-cutting sequence Hadamard encoding (CC-Hadamard), in which basis patterns are ordered according to decreasing optical coherence area to optimize the light throughput \cite{Yu2019Sensors}. As shown in Figs. \ref{fig3}(c1-c3), the Hadamard encoding consistently yields superior SNR and higher image fidelity across all tested resolutions when compared with both raster and random encoding strategies. In the above measurements, the MIR illumination power is fixed at 25 pJ/pulse, and the integration time for each pattern is set to be 0.1 s. The reconstruction performances for other objects with letters ``C", ``N", and ``U" are presented in Figs. \ref{fig3}(d1-d3), (e1-e3), and (f1-f3), respectively.

For the image reconstruction, we employ a total variation minimization algorithm (TVAL3), which offers rapid convergence, high reconstruction accuracy, and compatibility with diverse constraint models \cite{Li2013Comput}. To further improve image quality and suppress unwanted background, a differential measurement scheme is employed, where the signal is calculated as the intensity difference between each encoding pattern and its inverse \cite{Edgar2019NP, Wang2023NC}. This approach effectively cancels static offsets induced by the degenerate two-photon absorption (D-2PA) from the 1550 nm pump and mitigates low-frequency fluctuations in the illumination, thereby enhancing the SNR, particularly under low-light conditions. While higher pump power can enhance the ND-2PA signal through its bilinear intensity dependence, the pump level is carefully optimized to avoid excessive D-2PA background, detector saturation, and potential optical distortions in the projection path. This trade-off ensures stable system operation and high-fidelity spatial modulation throughout the imaging process. In our imaging system, the minimum pixel size of the optical mask is 7 $\mu$m, which is smaller than the achieved MIR imaging resolution of approximately 11 $\mu$m, as determined using a USAF 1951 resolution test target (see Supplementary Note 3). Owing to the shorter NIR wavelength used for spatial modulation, the encoding resolution can be further improved to the micrometer scale by employing a high-numerical-aperture objective lens, which significantly surpasses the pixel dimensions of conventional infrared cameras. This high-resolution optical masking enables diffraction-limited imaging performance in the MIR spectral range.

\subsection*{High-sensitivity compressive imaging}

\begin{figure*}[t!]
	\centering
	\includegraphics[width=0.9\textwidth]{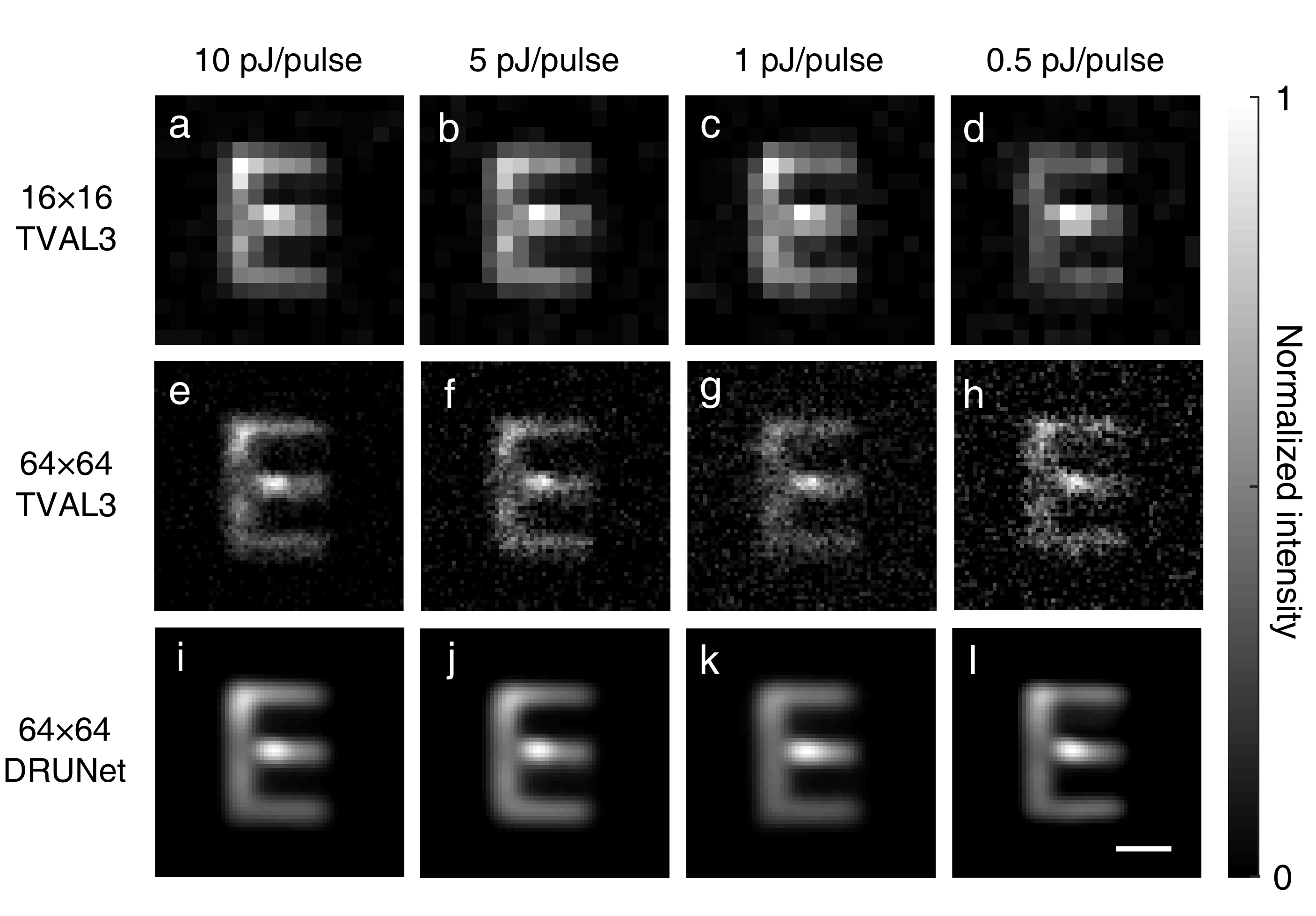}		
	\caption{Imaging performances under various illumination powers with two reconstruction algorithms. (a-d, e-h) Reconstructed images with 16$\times$16 (a-d) and 64$\times$64 (e-h) pixels by using the TVAL3 algorithm for incident MIR powers from 10 to 0.5 pJ/pulse. The integration time for each pattern is adapted to maintain the total number of recorded photons. (i-l) Recovered images with 64$\times$64 pixels using the DRUNet deep learning method. The scale bar denotes 100 $\mu$m.}
	\label{fig4}
\end{figure*}

In the following, we evaluate the detection sensitivity of the MIR single-pixel imaging system. Comparing to the TPA imaging based on focal plane arrays \cite{Knez2020Light, Liu2022IEEE, Knez2022SA, Fang2021IEEE}, the proposed single-pixel approach offers enhanced detection sensitivity and improved noise resilience, particularly under photon-limited conditions \cite{Meng2024LSA}. Figure \ref{fig4} presents reconstructed images with resolutions of 16$\times$16 and 64$\times$64 pixels under varying illumination powers. To compensate for the reduced photon flux at lower pulse energies, the integration time per pattern is proportionally increased, allowing sufficient signal for reconstruction. At relatively high incident energies ($\textit{e.g.},$ 10 pJ/pulse), high-SNR images are obtained with short integration time (50 ms/pattern). As the pulse energy decreases ($\textit{e.g.},$ 1 pJ/pulse), extending the integration time (500 ms/pattern) partially compensates for SNR degradation. However, at the system's sensitivity limit ($\sim$0.5 pJ/pulse, see Supplementary Note 2), the ND-2PA signal becomes indistinguishable from background noise. Increasing the integration time beyond 1 s/pattern further introduces dark noise, which ultimately limits reconstruction quality. This degradation is particularly evident at higher resolutions and lower light levels, as shown in Fig. \ref{fig4}(h). To mitigate these limitations, we integrate a machine learning-based denoising strategy into the reconstruction pipeline. Specifically, the DRUNet algorithm, which is a convolutional neural network (CNN) designed to capture both local structures and global context, is employed to suppress image noise and enhance detail preservation \cite{Zhang2021IEEE}. As shown in Figs. \ref{fig4}(i-l), the algorithm substantially improves SNR and image quality under low-light conditions. It is worth noting that the localized intensity variations observed in the reconstructed images may stem from several factors beyond the intrinsic detector response. Minor spatial non-uniformities in the projected NIR pump patterns can introduce encoding artifacts, for instance due to DMD surface imperfections, projection aberrations, or beam inhomogeneity. In addition, the use of deep learning reconstruction ($\textit{e.g.},$ DRUNet) under photon-starved or compressed sampling conditions can sometimes amplify localized features, producing apparent brightness enhancements if not fully regularized.

\begin{figure}[t!]
	\centering
	\includegraphics[width=0.6\columnwidth]{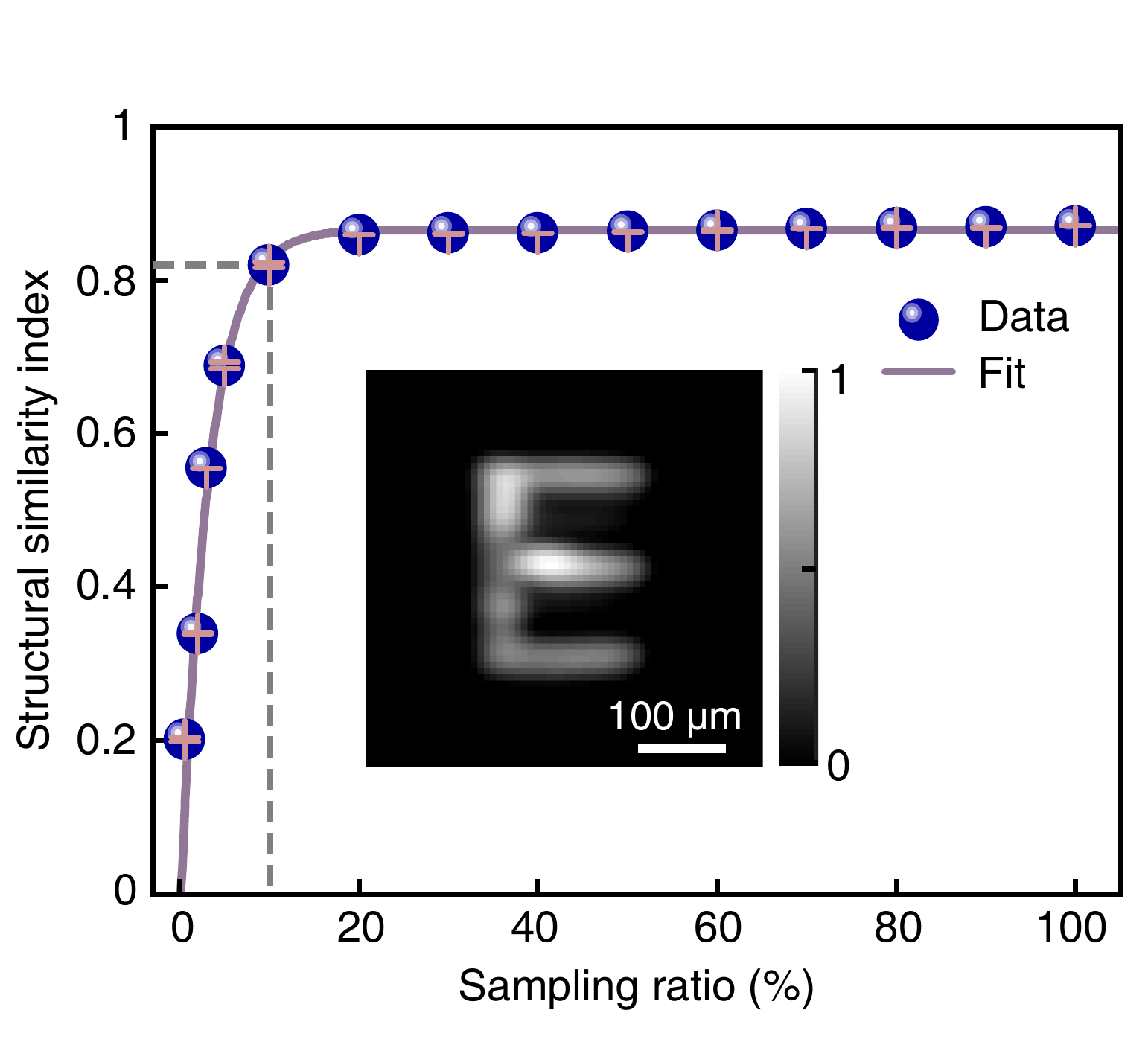}
	\caption{MIR compressive imaging performance. Structural similarity index (SSIM) of reconstructed images versus sampling ratio under a fixed incident MIR power of 0.5 pJ/pulse. The images are reconstructed using the DRUNet algorithm. The solid line is an exponential function fit. Inset shows the reconstructed image at a sampling ratio of 10$\%$.}
	\label{fig5}
\end{figure} 

We further evaluate the performance of under-sampled image reconstruction in low-light conditions. Traditional imaging systems are constrained by the Shannon-Nyquist sampling theorem, which dictates that the sampling rate must be at least twice the highest spatial frequency present in the signal. Consequently, achieving higher image resolutions typically requires a larger number of detector elements, increasing system complexity and cost. In contrast, single-pixel imaging, when combined with optimized encoding matrices and compressed sensing algorithms, enables accurate image reconstruction from significantly fewer measurements than the total number of image pixels \cite{Duarte2008IEEE}. In our experiment, under MIR illumination at 0.5 pJ/pulse, we randomly selected $M$ rows from the CC-Hadamard matrix to generate projection patterns. The total intensity recorded by the detector for each pattern is used to reconstruct an $N$-pixel image using the DRUNet algorithm. The sampling ratio is defined as $M/N$. To quantitatively assess reconstruction quality, we employ the structural similarity index (SSIM). As shown in Fig. \ref{fig5}, the SSIM remains above 80\% even at a sampling ratio as low as 10\%, with a representative reconstructed image shown in the inset. These results demonstrate that our system achieves high-sensitivity MIR imaging with up to 90\% compression under low-light conditions, substantially reducing both acquisition time and data storage requirements compared to conventional methods.

\subsection*{Broadband MIR multispectral imaging}
\begin{figure*}[t!]
	\centering
	\includegraphics[width=0.9\textwidth]{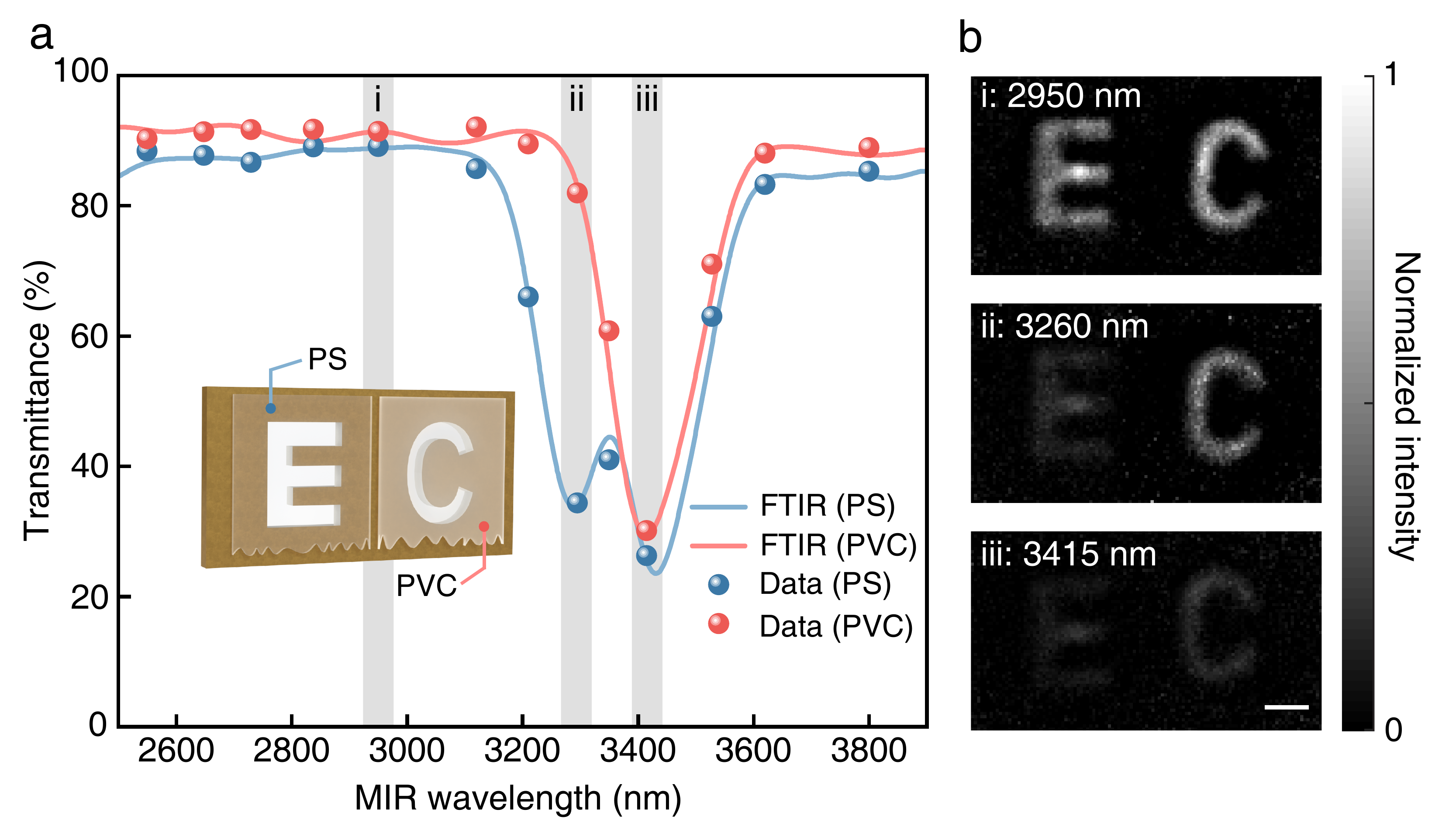}
	\caption{MIR spectral imaging to chemical identify thin-film samples. (a) Measured transmission spectra  for polystyrene (PS) and polyvinyl chloride (PVC) films, which are referenced to the FTIR traces with a spectral resolution of 90 cm$^{-1}$. (b) Monochromatic images for three selective spectral bands at 2950 nm (i), 3260 nm (ii), and 3415 nm (iii), respectively. The scale bar represents 100 $\mu$m.}
	\label{fig6}
\end{figure*}

Finally, we demonstrate the system's capability for chemically selective imaging of samples with distinct compositions and morphological features. Polystyrene (PS) and polyvinyl chloride (PVC) thin films are chosen as test samples and mounted on a transmission mask patterned with letters, as shown in the inset of Fig. \ref{fig6}. The films are sequentially positioned within the imaging field of view, and their transmittance spectra are recorded across the 2500-3800 nm MIR tuning range. The measured spectra show good agreement with reference data from Fourier-transform infrared (FTIR) spectroscopy. Three characteristic wavelengths (2950 nm, 3260 nm, and 3415 nm) are selected for single-pixel imaging. To enhance image contrast, intensity values are normalized across all reconstructed images. As shown in Fig. \ref{fig6}, the image at 3260 nm clearly differentiates the two materials based on their distinct absorption features. These results validate the system's potential for chemically specific detection. Additional multispectral imaging results across the full MIR range are provided in Supplementary Note 4.

Note that MIR imaging uniquely combines label-free chemical contrast with improved subsurface penetration due to reduced medium scattering. These features are highly desirable for biomedical diagnostics \cite{Shi2020NM} and material analysis \cite{Israelsen2019LSA}, yet their practical implementation remains constrained by the lack of sensitive, broadband MIR detection technologies. The imaging architecture developed in this work could directly address this challenge by enabling high-sensitivity, room-temperature MIR detection with wide spectral coverage and sub-pJ sensitivity, thus laying a technological foundation for emerging applications in MIR spectroscopy, deep-tissue imaging, and molecular diagnostics \cite{Knez2022SA}.

\section*{Conclusion}

In conclusion, we propose and experimentally demonstrate a MIR single-pixel TPA imaging scheme based on nonlinear structured pumping, which leverages a spatially unresolved silicon single-photon detector to achieve high-sensitivity, high-resolution, and wide-field imaging. The core innovation lies in exploiting the ND-2PA effect to effectively transfer the spatial encoding from the NIR pump beam to the MIR domain, thereby establishing a structured detection mechanism via nonlinear optical modulation. Unlike direct MIR modulation approaches, this method circumvents long-wavelength diffraction-induced distortions and enables high-fidelity spatial encoding, laying the foundation for high-resolution MIR imaging. Compared to conventional point-scanning TPA-based MIR imaging, the integration of dynamic DMD-based modulation in our system overcomes the trade-off between spatial resolution and acquisition speed. When combined with compressive sensing algorithms, the sampling efficiency can be substantially improved. Moreover, in contrast to array-based TPA imaging systems, the single-pixel architecture retains a wide field of view, enhances the SNR by collecting incident light into a single-element sensor, and significantly reduces system complexity and manufacturing cost.

To go beyond the achieved imaging performances, there exist several aspects that warrant further exploration. First, the spatial sampling resolution of the all-optical masking can be enhanced by resorting to objective lenses with high numerical apertures, enabling finer projection of DMD patterns onto the detector plane. This allows more encoding elements to be packed within the same detector area. Moreover, the total number of resolvable elements could be significantly increased, potentially reaching the megapixel level, by combining this approach with a large-area photodetector such as a photomultiplier tube (PMT). Second, the spectral window of the ND-2PA imager could be extended to cover longer infrared wavelengths over 8 $\mu$m by adopting a pump wavelength around 1.3 $\mu$m. Notably, the detector materials can be replaced with direct bandgap semiconductors such as InGaAs or GaN \cite{Fishman2011NP, Knez2022SA}, which offer enhanced TPA efficiency and support improved sensitivity over a broad spectral coverage. Third, the synchronized pulsed optical gating enables precise control over the temporal overlap between the MIR and pump pulses, making the proposed imaging system inherently capable of extracting axial depth information. This opens the possibility for three-dimensional infrared imaging with high transverse and axial resolution \cite{Sun2016NC}, potentially enabling a wide range of applications, including biological tissue tomography \cite{Shi2020NM, Feng2021LSA} and defect inspection in semiconductor materials \cite{Israelsen2019LSA, Liu2021LSA}. 
\newline

\backmatter

\bmhead{Supplementary information}
Supporting Information is accompanied to present more experimental details. 

\bmhead{Author's contribution}
K.H., H.M., and H.Z. conceived the project and designed the experiments. H.M. and K.H. built the system, performed experiments, and processed data. Y.L. built detector. Z.H. analyzed the imaging data. H.M. and K.H. wrote the
manuscript draft. All authors were involved in discussions and contributed to the manuscript editing.

\bmhead{Funding}
Shanghai Pilot Program for Basic Research (TQ20220104), National Natural Science Foundation of China (62175064, 62235019, 62035005), Innovation Program for Quantum Science and Technology (2023ZD0301000), Shanghai Municipal Science and Technology Major Project (2019SHZDZX01); Natural Science Foundation of Chongqing (CSTB2023NSCQ-JQX0011, CSTB2022TIAD-DEX0036), China Post doctoral Science Foundation (2024M760918, 2025T180224).

\bmhead{Availability of data and materials}
The data that support the findings of this study are available from the corresponding author upon reasonable request.

\section*{Declarations}
\bmhead{Ethics approval and consent to participate}
There is no ethics issue for this paper.

\bmhead{Consent for publication}
All authors agreed to publish this paper.

\bmhead{Conflict of Interest}
The authors declare that they have no conflict of interests.

\end{document}